\documentclass[aps,prl,reprint,superscriptaddress,longbibliography]{revtex4-1}
\usepackage{graphicx}
\usepackage{latexsym}
\usepackage{amsmath}
\usepackage{txfonts}
\usepackage{helvet}
\usepackage{braket}
\usepackage{xcolor}
\usepackage[%
  colorlinks=true,
  urlcolor=blue,
  linkcolor=blue,
  citecolor=blue
]{hyperref}

\begin{document}
\title{A molecular diamond lattice antiferromagnet as a Dirac semimetal candidate} 

\author{Yasuhiro Shimizu}
\affiliation{Department of Physics, Nagoya University, Furo-cho, Chikusa-ku, Nagoya 464-8602, Japan.}
\author{Akihiro Otsuka}
\affiliation{Division of Chemistry, Graduate School of Science, Kyoto University, Sakyo-ku, Kyoto 606-8502, Japan.}
\author{Mitsuhiko Maesato}
\affiliation{Division of Chemistry, Graduate School of Science, Kyoto University, Sakyo-ku, Kyoto 606-8502, Japan.}
\author{Masahisa Tsuchiizu}
\affiliation{Department of Physics, Nara Women's University, Kitauoyanishi-machi, Nara 630-8506, Japan.}
\author{Akiko Nakao}
\affiliation{Neutron Science and Technology Center, CROSS Tokai, Ibaraki 319-1106, Japan.}
\author{Hideki Yamochi}
\affiliation{Division of Chemistry, Graduate School of Science, Kyoto University, Sakyo-ku, Kyoto 606-8502, Japan.}
\author{Takaaki Hiramatsu}
\affiliation{Faculty of Agriculture, Meijo University, Shiogamaguchi 1-501 Tempaku-ku, Nagoya 468-8502, Japan.}
\author{Yukihiro Yoshida}
\affiliation{Division of Chemistry, Graduate School of Science, Kyoto University, Sakyo-ku, Kyoto 606-8502, Japan.}
\affiliation{Faculty of Agriculture, Meijo University, Shiogamaguchi 1-501 Tempaku-ku, Nagoya 468-8502, Japan.}
\author{Gunzi Saito}
\affiliation{Faculty of Agriculture, Meijo University, Shiogamaguchi 1-501 Tempaku-ku, Nagoya 468-8502, Japan.}
\affiliation{Toyota Physical and Chemical Research Institute, Nagakute, 480-1192, Japan.}

\begin{abstract}
The ground state of a molecular diamond-lattice compound (ET)Ag$_4$(CN)$_5$ is investigated by the magnetization and nuclear magnetic resonance spectroscopy. We found that the system exhibits antiferromagnetic long-range ordering with weak ferromagnetism at a high temperature of 102 K owing to the strong electron correlation. The spin susceptibility is well fitted into the diamond-lattice Heisenberg model with a nearest neighbor exchange coupling of 230 K, indicating the less frustrated interactions. The transition temperature elevates up to $\sim$195 K by applying pressure of 2 GPa, which records the highest temperature among organic molecular magnets. The first-principles band calculation suggests that the system is accessible to a three-dimensional topological semimetal with nodal Dirac lines, which has been extensively searched for a half-filling diamond lattice. 

\end{abstract}

\maketitle

A half-filling diamond lattice has been recently attracted great interests as an example of three-dimensional (3D) Dirac semimetal with the linearly-crossing band dispersion near the Fermi level \cite{Fu, Zhang, Young, Ryu} along with the appealing example including Na$_3$Bi and Cd$_3$As$_2$ \cite{Liu, Cava}. The system can be a strong topological insulator in the presence of spin-orbit coupling as a 3D analogue to graphene \cite{Fu}. Despite the popular crystal structure, the material with the half-filled band has been known only in a putative material BiO$_2$ \cite{Young}. In the counterpart insulating system, the frustrated local moment on the diamond lattice has been extensively studied as a spin liquid candidate \cite{Bergman, Bernier, Buessen}. A typical example is the magnetic spinel ($AB_2C_4$) with the $A$ site diamond lattice \cite{Fritsch, Krimmel, Plumb, Ge, Chamorro, Iakovleva, MacDougall, MacDougall2, Zaharko}, where the properties of the disordered state are under intense debate. A (topological) Mott transition is expected to occur from a spin disordered phase to a Dirac semimetal phase by tuning the electron correlation \cite{Zhang, Bergman}.

Organic molecular compounds have provided the platform for investigating the pressure-tuned Mott transition for the soft crystal. The well-studied Mott-Hubbard systems such as $\kappa$-(ET)$_2$X and Z[Pd(dmit)$_2$]$_2$ possess a quasi-two-dimensional triangular lattice of the molecular dimer unit \cite{Shimizu, Itou, Isono} owing to the anisotropic intermolecular interactions between planar molecules. Thus there are only a few example of 3D molecular compounds including the diamond lattice, except for the inorganic-organic hybrid system such as Li(TCNE) and Cu(DCNQI)$_2$ \cite{Miller, Miller2, Ermer}, and no example is known for the half-filling diamond lattice consisting of organic molecules. The material search for 3D molecular compounds would be important for giving high-temperature magnets and superconductors. 

\begin{figure}
\begin{center}
	\includegraphics[scale=0.68]{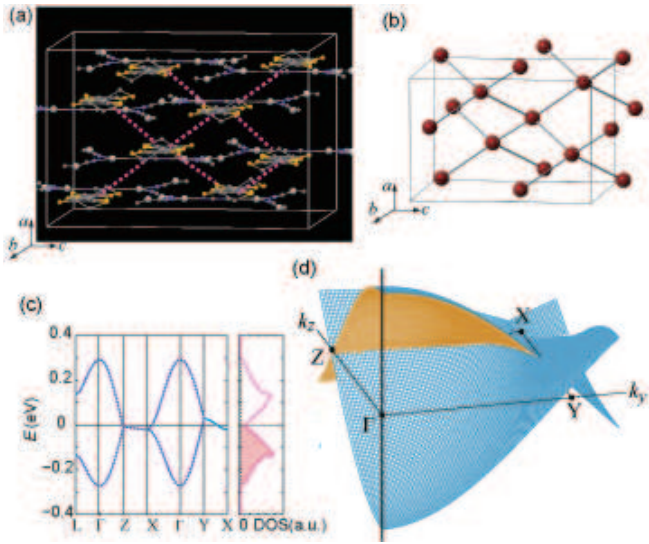}
	\caption{\label{Fig1} 
	(a) Crystal structure of (ET)Ag$_4$(CN)$_5$ with a diamond lattice. ET molecules are deviated by [$\pm 1/4$, $\pm 1/4$, $\pm 1/4$] each other in the unit cell. Each molecule has four equivalent nearest neighbor interactions (dotted lines). (b) Diamond lattice of ET where the red sphere represents the centroid of ET. (c) Highest-occupied molecular orbital (HOMO) bands based on the first-principles calculation. The degenerate band crosses the Fermi level along $Z-X-Y$ due to second neighbor transfers, which gives electron and hole-like Fermi surfaces [Fig. S2(c)]. Dotted lines are the fitting results with tight-binding parameters of transfer integrals \cite{SI}. The density of states (DOS) has the V-shaped energy dependence centered at the Fermi energy. (d) 3D projection of the band dispersion, showing nodal Dirac lines along the symmetry points along $Z-X-Y$. 
	}
	\end{center}
	\end{figure}

We present here the molecular material (ET)Ag$_4$(CN)$_5$ \cite{Geiser} as a prime example of the 3D diamond lattice. It possesses the extremely high-symmetry crystal structure of the orthorhombic $Fddd$ lattice with the lattice constants: $a = 13.215(9)$ \AA, $b = 19.4783(1)$ \AA, and $c = 19.6506(1)$ \AA. Each monovalent ET molecule is surrounded by the honeycomb framework of the closed-shell polyanion [Ag$_4$(CN)$_5$]$_{\infty }^-$ in the $bc$ plane, as shown in Fig. \ref{Fig1}(a). The next stacking layer along the $a$ axis is deviated by [$1/4$, $1/4$, $1/4$] in the unit cell. This unique packing pattern is distinct from the layered or columnar structure in typical molecular solids. The alternating stacking makes ET molecules to construct a diamond lattice with four equivalent nearest neighbor transfers. Therefore, the system is regarded as a half-filling diamond lattice of the molecular unit [Fig. 1(b)], which can be either a spin-$1/2$ Heisenberg antiferromagnet or a 3D Dirac semimetal, depending on the strength of electron correlations. 

In this Letter, we investigate the ground state of the diamond-lattice molecular compound (ET)Ag$_4$(CN)$_5$ through the resistivity, magnetization, and nuclear magnetic resonance (NMR) measurements. As expected for the half-filling organics composed of the monovalent ions, the ground state behaves as a Mott insulator with a spin-$1/2$ on each molecule. We determined the spin structure and dynamics through the angular dependence of $^{13}$C NMR at ambient pressure. Together with the band calculation, we discuss the possible Dirac semimetal phase emerging from the antiferromagnetically ordered phase. 

Single crystals of (ET)Ag$_4$(CN)$_5$ were prepared by galvanostatic electrooxidation of ET in a 1,1,2-trichloroethane solution of KAg(CN)$_2$ and 18-crown-6 ether. The obtained rhombohedral shaped crystals of (ET)Ag$_4$(CN)$_5$ were carefully separated from minor co-products of $\kappa$-(ET)$_2$Ag$_2$(CN)$_3$ \cite{Shimizu2}, $\alpha^\prime$-(ET)$_2$Ag(CN)$_2$ \cite{Beno}, and $\kappa$-(ET)$_2$Ag(CN)$_2$ $\cdot$ H$_2$O \cite{Kurmoo}. Resistivity was measured with a four-probe dc method at ambient and hydrostatic pressures, where gold wires were attached to a single crystal using the carbon paint. Hydrostatic pressure was applied using a BeCu piston cylinder cell with Daphne 7373 oil and monitored by the manganin wire resistance. Magnetic susceptibility was measured for a polycrystalline sample by a superconducting quantum interference device magnetometer (Quantum Design MPMS-XL). The core diamagnetism value was calculated as a sum of Pascal's constants ($-3.74 \times 10^{-4}$ emu mol$^{-1}$). $^1$H and $^{13}$C NMR spectra were obtained for a single crystal in a static magnetic field of 2.0 T and 8.5 T, respectively, which were calibrated using the resonance frequency of the standard sample, tetramethylsilane (TMS). 

The band structure of (ET)Ag$_4$(CN)$_5$ was obtained from the first-principles density-functional-theory (DFT) calculation based on the generalized gradient approximation with the WIEN2K code, as shown in Fig. 1(c). Here the band structure was evaluated without considering the anion having the orientational CN/NC disorder (Fig. S2) \cite{SI}. The Fermi energy is located at the half-filling position, where two HOMO bands cross. The two bands are degenerate along $Z-X$ and $Y-X$ directions, characteristic in the diamond lattice \cite{Chadi}. The result qualitatively agrees with the extended H${\rm \ddot{u}}$ckel and the tight-binding calculation \cite{Geiser, Geiser2}. Along the symmetry positions of the band, nodal Dirac lines appear near the Fermi level, as shown in Fig. \ref{Fig1}(d), consistent with the result for the cubic diamond lattice \cite{Young}. 

\begin{figure}
\begin{center}
	\includegraphics[scale=0.52]{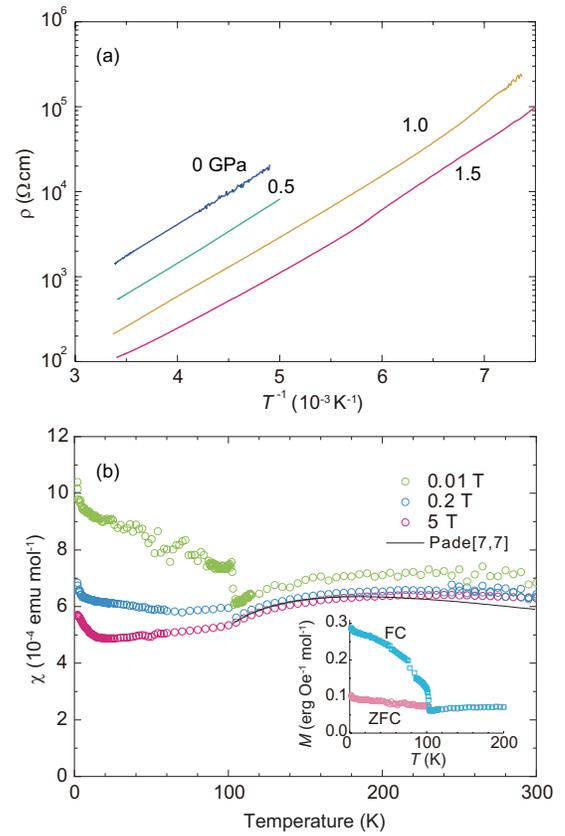}
	\caption{\label{Fig2} 
	(a) Inverse temperature dependence of resistivity $\rho$ measured along the $b$ axis of (ET)Ag$_4$(CN)$_5$ at 0.0, 0.5, 1.0, and 1.5 GPa. (b) Temperature dependence of magnetic susceptibility $\chi$ for zero-field-cooling (ZFC) at $H$ = 0.01, 0.2, and 5.0 T, where the core diamagnetic contribution was subtracted. The solid curve is a fitting result using a diamond-lattice Heisenberg model \cite{Oitmaa} with the Pad${\rm \acute{e}}$ approximant [7, 7] and $J = 230$ K. Inset: ZFC and FC magnetization at 0.01 T. 
	}
\end{center}
\end{figure}

Despite the semimetallic band structure, the resistivity $\rho$ exhibits insulating temperature ($T$) dependence, as shown in Fig. \ref{Fig2}(a). The activation energy is obtained as $E_b$ = 0.15 eV along the $b$ axis at ambient pressure. Together with the paramagnetic spin susceptibility $\chi$ [Fig. \ref{Fig2}(b)], the system is regarded as a Mott insulator due to the on-site Coulomb interaction $U (\sim 1$ eV for the monovalent ET) greater than the bandwidth ($W \sim 0.57$ eV) [Fig. 1(c)]. By applying hydrostatic pressure, $\rho$ is suppressed by an order of magnitude, and $E_b$ decreases to 0.12 eV at 1.5 GPa. 

The temperature dependence of $\chi$ [Fig. \ref{Fig2}(b)] is distinct from the Curie-Weiss law in classical paramagnets, but exhibits a broad maximum around 220 K. The weak temperature dependence of $\chi$ is similar to the triangular-lattice anitiferromagnet such as $\kappa$-(ET)$_2$Cu$_2$(CN)$_3$ \cite{Shimizu}, and the 1D Heisenberg antiferromagnet such as Sr$_2$CuO$_3$ \cite{Motoyama}, suggesting significant quantum fluctuations. The experimental data are well fitted by the high-temperature series expansion of the $S = 1/2$ diamond-lattice Heisenberg model with the Pad${\rm \acute{e}}$ approximant [7, 7] \cite{Oitmaa}, yielding the antiferromagnetic exchange coupling $J = 230\pm 10$ K. An indication of the magnetic transition is observed at $T_{\rm N}$ = 102 K, where $\chi$ shows an abrupt increase at low fields. The prominent magnetic field ($H$) dependence in the magnetization $M$ and the thermal hysteresis [Fig. \ref{Fig2}(b) inset, Fig. S3] highlight weak-ferromagnetism due to the canting of antiferromagnetic moments. 

	\begin{figure}
	\includegraphics[scale=0.75]{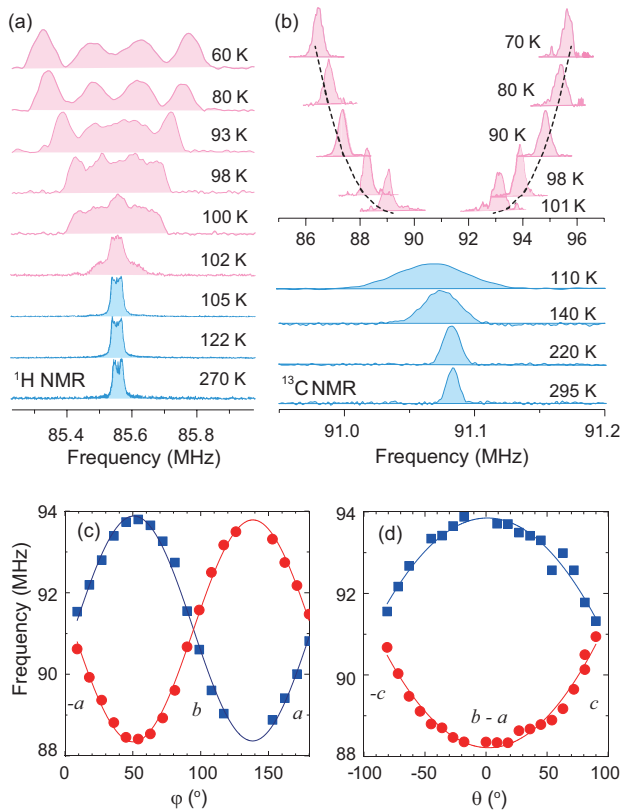}
	\caption{\label{Fig3} 
	(a) $^1$H NMR spectrum along the $a$ axis at 2.0 T. (b) $^{13}$C NMR spectrum at 8.5 T applied along [110] corresponding to the magic angle where the $^{13}$C-$^{13}$C dipole coupling vanishes for $T > T_{\rm N}$. Dotted curves represent temperature dependence of order parameters with a critical exponent $\beta = 0.36$, as expected in 3D Heisenberg antiferromagnets. The vertical axis scales to temperature. (c, d) Angle dependence of $^{13}$C NMR frequency in the ordered state at 95 K, where $\varphi $ and $\theta $ are defined by the angles measured from the $-a$ and $-a + b$ directions toward the $b$ and $c$ axes, respectively.  
	}
	\end{figure}

Microscopic evidence for long-range magnetic ordering is given by $^1$H and $^{13}$C NMR measurements on the single crystal of (ET)Ag$_4$(CN)$_5$, as shown in Fig. \ref{Fig3}. Above 105 K, the $^1$H NMR spectrum [Fig. \ref{Fig3}(a)] represents the $^1$H-$^1$H nuclear dipole coupling with the $T$-independent linewidth ($\sim 50$ kHz). The spectrum begins to split into four below 102 K. The splitting develops upon cooling temperature, signaling the emergence of huge local fields parallel and antiparallel to the external field. The presence of two inequivalent $^1$H sites on an ET molecule further splits each spectrum into two.  

To determine the ordered spin texture, the $^{13}$C NMR spectrum was measured for the isotope enriched crystal at the central double-bonded C sites with the high electron density \cite{Larsen}. There is only a single $^{13}$C site manifested as a sharp line in the paramagnetic state. The linewidth grows as spin fluctuations slow down toward $T_{\rm N}$. Below $T_{\rm N}$, the spectrum splits into two due to the antiferromagnetic order, consistent with $^1$H NMR. The angular dependence of the NMR frequency $\omega$ around the crystal axes [Fig. \ref{Fig3}(c,d)] shows that the local field ${\bf H}_{\rm loc}$ exhibits a minimum or maximum against the external magnetic field ${\bf H}_0$ parallel to ${\bf b \pm a}$. A rotation of ${\bf H}_0$ from the ${\bf b-a}$ to ${\bf c}$ confirms ${\bf H}_{\rm loc} \parallel  {\bf b \pm a}$ [Fig. \ref{Fig3}(d)]. 

Here the $^{13}$C nuclear spin experiences a sum of the external field ${\bf H}_0$ and the spontaneous local field ${\bf H}_{\rm loc}$ produced by the magnetic moment ${\bf M}_{\rm loc}$: $\omega$ is given by $\omega = \gamma_n H_{\rm eff} = \gamma_n (|{\bf H}_0 + {\bf H}_{\rm loc}|) = \gamma_n \sqrt{H_0^2 + H_n^2 + 2H_0 H_n {\rm cos}\vartheta }$, where $\gamma_n$ is the $^{13}$C nuclear gyromagnetic ratio ($10.7054$ MHz/T) and $\vartheta $ is the angle between ${\bf H}_0$ and ${\bf H}_{\rm loc}$. ${\bf H}_{\rm loc}$ is given by ${\sf A} {\bf M}_{\rm loc} = (A_{aa}M_a, A_{bb}M_b, A_{cc}M_c)$ using the hyperfine coupling tensor ${\sf A}$ with diagonal components $A_{\alpha \alpha} (\alpha=a, b,c)$ and ${\bf M}_{\rm loc}  = (M_a, M_b, M_c)$. We determined ${\sf A}$ from the $K-\chi$ plot (Fig. S5) as $(A_{aa}, A_{bb}, A_{cc}) = (-0.21. -0.21, 1.64)$ T/$\mu_{\rm B}$, which yields the isotropic Fermi contact ($\alpha = 0.41$ T/$\mu_{\rm B}$) and the anisotropic dipole hyperfine coupling ($\beta = 0.62$ T/$\mu_{\rm B}$) as expected for the $sp_2$ orbital \cite{Miyagawa2}.

\begin{figure}
	\begin{center}
	\includegraphics[scale=0.62]{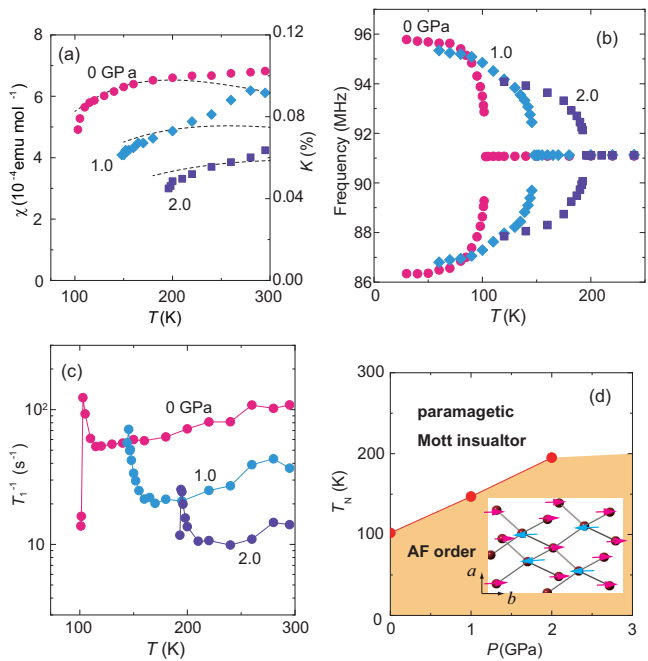}
	\caption{\label{Fig4} 
	(a) Static spin susceptibility $\chi$ obtained from $^{13}$C Knight shift $K = A\chi/(N\mu_{\rm B})$ with the hyperfine coupling constant $A=0.83 {\rm T}/\mu_{\rm B}$ determined at 0 GPa. The magnetic field was applied parallel to [110]. Solid curves are the series expansion of the diamond-lattice Heisenberg model \cite{Oitmaa}. (b) $^{13}$C NMR frequency defined by the spectral peak across $T_{\rm N}$ at 0, 1.0, and 2.0 GPa, where the magnetic field is applied along [110]. (c) The nuclear spin-lattice relaxation rate $T_1^{-1}$. The steep increase toward magnetic ordering occurs due to slowing down to spin fluctuations. (d) Pressure-temperature phase diagram of (ET)Ag$_4$(CN)$_5$ based on the NMR measurements. Inset shows the spin configuration determined by $^{13}$C NMR.
	}
 \end{center}
	\end{figure}

The temperature dependence of $^{13}$C NMR spectrum for ${\bf H}_0 \parallel [110]$ [Fig. \ref{Fig3}(b)] demonstrates the evolution of the order parameter, following a scaling law $\sim (T_{\rm N} -T)^\beta$ with a critical exponent $\beta = 0.36 \pm 0.02$ for $T > 70$ K [Fig. \ref{Fig3}(b)]. It is in good agreement with $\beta$ = 0.368 in the 3D Heisenberg model \cite{Benner}. By using the obtained ${\sf A}$, we determined the magnetic moment ${\bf M}_{\rm loc} = (0.114, 0.889, 0.0)\mu_{\rm B}$ with the magnitude $|{\bf M}_{\rm loc}| = 0.90 \pm 0.02\mu_{\rm B}$ ($\mu_{\rm B}$: the Bohr magneton) at 40 K. Namely, the easy axis of the moment is directed close to the $b$ axis with the collinear configuration, as schematically shown in Fig. \ref{Fig4}(d), consistent with the theoretical ground state for the less frustrated diamond lattice \cite{Bergman, Bernier}. The tiny canting of the moment ($\sim 0.012^\circ$ obtained from the $M-H$ curve, Fig. S3) was not detected within the accuracy of the NMR measurement. 

Interamolecular interactions can be sensitively tuned for the soft organic crystal by applying pressure. Figure 4 shows the $^{13}$C Knight shift $K$ and the nuclear spin-lattice relaxation rate $T_1^{-1}$ for (ET)Ag$_4$(CN)$_5$ under hydrostatic pressure up to 2 GPa. Here $K$ is converted into the local spin susceptibility $\chi$ using a relation $K = A\chi/(N\mu_{\rm B})$, $N$: the Avogadro number) with the hyperfine coupling constant $A$ at ambient pressure [Fig. 4(a)], whereas $T_1^{-1}$ measures the dynamical spin susceptibility that scales to $J^{-1}$ at high temperatures: $T_1^{-1} = \sqrt{\frac{\pi}{3}}\frac{A^2\sqrt{S(S+1)}}{\hbar J\sqrt{z}}$ \cite{Moriya}. We obtained $J$ = 240$\pm 20$ K at 0 GPa, in agreement with that obtained from $\chi$. By applying pressure, $\chi$ is suppressed owing to an increase in $J$. Fitting of $\chi$ into the diamond-lattice Heisenberg model \cite{Oitmaa} allows a rough estimate of the exchange coupling: $J = 310 \pm 40$ K at 1.0 GPa and $400 \pm 30$ K at 2.0 GPa. 

Simultaneously, $T_{\rm N}$ obtained from the spectral splitting [Fig. 4(b)] and the sharp $T_1^{-1}$ peak [Fig. 4(c)] elevates as we increase the pressure: $T_{\rm N} = 150$ K at 1.0 GPa and 195 K at 2.0 GPa [Fig. 4(d)]. It corresponds to the highest magnetic transition temperature among molecular materials including the organic charge-transfer salt \cite{Smith, Miyagawa2}, the transition-metal hybrid system such as [Au(tmdt)$_2$] ($T_{\rm N} \sim 110$ K) \cite{Hara}, and the C$_{60}$ complex such as (NH$_3$)KRb$_2$C$_{60}$ ($T_{\rm N} = 76$ K) \cite{Takenobu}. In contrast to the mean-field theory giving $T_{\rm N} \sim J = \Theta$ ($\Theta$: Weiss temperature) for the diamond lattice, the experimentally obtained $T_{\rm N}$ is suppressed to the temperature scale of $\sim J/2$. It is consistent with the significant quantum fluctuations in the $S = 1/2$ Heisenberg antiferromagnet \cite{Oitmaa}. Despite an increase of $T_{\rm N}$, the magnetic moment is suppressed upon increasing pressure [Fig. \ref{Fig4}(b)]. The moment contraction is attributable to quantum fluctuations due to the electron itinerancy and a weak dimerization of ET molecules. 


Theoretically, the diamond lattice involves geometrical frustration due to 12 next nearest neighbor interactions $J^\prime$. An introduction of the small $J^\prime \sim J/8$ can suppress $T_{\rm N}$ and induce a spin liquid state \cite{Bergman, Bernier}. The frustration is released by strong thermal or quantum fluctuations via order-by-disorder mechanism, where the ground state is determined by an entropical or energetical selection \cite{Bergman, Bernier}. In the present case, however, a tight-binding calculation suggests the negligible $J^\prime/J \sim (t^\prime/t)^2 < 0.004$. Indeed, the obtained $T_{\rm N}/J = 0.46$ at 0 GPa and 0.49 at 2.0 GPa are consistent with the $S = 1/2$ diamond-lattice Heisenberg model including only the nearest neighbor interaction \cite{Oitmaa}. Furthermore, in the real system, the highly degenerated (six-fold) spin structure on the diamond lattice should be lifted by single-ion anisotropy (spin-orbit coupling) or lattice distortion (spin-phonon coupling), triggering the magnetic order. Whereas the $g$-value is nearly isotropic ($g_a = 2.0026$, $g_b = 2.0157$, $g_c = 2.0069$) in (ET)Ag$_4$(CN)$_5$, the spin-orbit coupling as well as the spin-phonon coupling, which leads to Dzyaloshinskii-Moriya interaction, may play a key role in the spin texture. Despite the presence of the structural CN/NC disorder in the anion, analogous to $\kappa$-(ET)$_2X_2$(CN)$_3$ ($X$ = Ag, Cu) without magnetic ordering \cite{Shimizu, Shimizu2}, the present system with a less frustrated diamond lattice exhibits the high $T_{\rm N}$. It suggests that the disorder potential from the counter ion plays an negligible effect in the magnetic ground state.

Our finding demonstrates the potential of the molecular conductors for three-dimensional and high-transition temperature magnets through the combination of polymeric counter ions. The high-$T_{\rm N}$ Mott insulator may host high-$T_{\rm c}$ superconductivity across the Mott transition under high pressure, because the energy scale of the exchange interaction ($>$ 400 K) for (ET)Ag$_4$(CN)$_5$ may be greater than those of the dimer ET salts such as $\beta^\prime$-(ET)$_2$ICl$_2$ ($T_c$ = 14.2 K at 8.2 GPa) \cite{Taniguchi} and C$_{60}$ complex such as Cs$_3$C$_{60}$ ($T_{\rm c}$ = 38 K) \cite{Takabayashi, Takenobu}. As shown by the band calculation, the metallic phase induced by high pressure may be 3D Dirac semimetal, which has a topological surface state in the presence of spin-orbit coupling and inversion symmetry breaking \cite{Fu, Young}. The emergence of the Dirac semimetal has recently observed in a single-component molecular conductor under high pressures \cite{Kato, Liu2018}.

To conclude, we investigated the ground state for the uniquely high-symmetry organic Mott insulator with the diamond lattice, (ET)Ag$_4$(CN)$_5$, which possesses nodal Dirac lines in the original band structure without electron correlations. Whereas the charge activation energy exceeds 0.1 eV at ambient pressure, the antiferromagnetic exchange interaction reaches $J$ = 230 K, and the long-range magnetic order with the weak ferromagnetism occurs at the high temperature of 102 K. Furthermore, the application of hydrostatic pressure enhances the transition temperature up to 195 K, which is highest among the molecular systems and thus anticipated to host the high-$T_c$ superconductivity at higher pressure.

This work was supported by JSPS KAKENHI Grant Numbers JP16K13836, JP17H05151, and JP16H04012, JP16H04139, and JP26288035.

\bibliographystyle{apsrev4-1} 
\bibliography{Bib_YS} 

\end{document}